\documentclass[prb,aps,amsfonts,eqsecnum,superscriptaddress,nofootinbib,longbibliography,notitlepage]{revtex4-1}
\usepackage{natbib}
\usepackage[utf8]{inputenc}
\usepackage{amsmath,amssymb,amsfonts,xcolor}

\newcommand{\ra}{\rightarrow}
\newcommand{\eps}{\epsilon}
\newcommand{\tE}{{\tilde E}}

\newcommand{\bb}{{\mathbf b}}
\newcommand{\rv}{{\bf r}}
\newcommand{\bD}{{\mathbf D}}
\newcommand{\bh}{{\mathbf h}}
\newcommand{\br}{{\mathbf r}}
\newcommand{\bB}{{\mathbf B}}
\newcommand{\bkappa}{{\mathbf\kappa}}
\newcommand{\RR}{{\mathbb R}}
\newcommand{\bX}{{\mathbf X}}
\newcommand{\bH}{{\mathbf H}}
\newcommand{\bq}{{\mathbf e}}
\newcommand{\bS}{{\mathbf S}}
\newcommand{\bJ}{{\mathbf J}}
\newcommand{\cD}{D}

\usepackage[colorlinks=true]{hyperref} 

\hypersetup{
    unicode=false,          
    pdftoolbar=true,        
    pdfmenubar=true,        
    pdffitwindow=false,     
    pdfstartview={FitH},    
    pdfsubject={},   
    pdfcreator={},   
    pdfproducer={}, 
    pdfkeywords={} {} {}, 
    pdfnewwindow=true,      
    colorlinks=true,       
    linkcolor=magenta, 
    citecolor=blue,        
    filecolor=magenta,      
    urlcolor=blue           
} 
\begin{document}

\title{Piezo-superconductivity: new effects in non-centrosymmetric superconductors}
\author{Anton Kapustin}
\email{kapustin@theory.caltech.edu}
\affiliation{California Institute of Technology, Pasadena, CA 91125, USA}
\author{Leo Radzihovsky}
\email{radzihov@colorado.edu}
\affiliation{Department of Physics and Center for Theory of Quantum Matter, University of Colorado, Boulder CO 80309, USA}

\begin{abstract}
We study  novel effects in non-centrosymmetric superconductors arising from their unique coupling of  Cooper-pair condensate and elasticity. We show that although the much discussed Lifshitz coupling is not observable in a  uniform bulk state, it strikingly endows dislocations with a fractional magnetic flux. We also predict a generation of voltage-free strain by a DC current in a $P$- and $T$- breaking Josephson junction. Viewing superconductors through the lens of higher-form symmetries we identify the Lifshitz coupling as a chemical potential for the  approximately conserved winding number, drawing an analogy  with pyroelectric insulators. 
\end{abstract}

\maketitle

\section{Introduction}
\subsection{Background and motivation}

Superconductors are one of the more thoroughly studied states of matter, exhibiting rich phenomenology, much of which is quite well mapped out and understood both from microscopic and effective field-theoretic points of view. Yet new phenomena are being discovered in materials that exhibit unconventional superconductivity. This includes such diverse materials as cuprate, iron-based, and heavy fermion superconductors.

In this paper we consider non-centrosymmetric  superconductors with broken time-reversal symmetry. We use equilibrium effective field-theory to explore their low-energy universal properties, including coupling to elastic degrees of freedom, that can and 
generally are neglected in the low-energy description of  centrosymmetric superconductors.
Among a number of predictions, we show that low-order current - elastic strain coupling unique to these materials leads to a nontrivial low-energy phenomenology that we refer to as piezo-superconductivity.


Non-centrosymmetric materials are those where the spatial inversion symmetry $P$ is broken. There are many classes of such materials, including ferromagnets and piezoelectric insulators. The study of non-centrosymmetric superconductors, however, has begun only relatively recently, see e.g., Refs. \onlinecite{Yip,Smidman2017} for a review. Superconductors  which break $P$ in the bulk are actually not that rare \cite{Yip}. $P$-breaking also occurs naturally in 2D films in the presence of spin-orbit interactions or as a surface effect. Microscopically, $P$-breaking implies that the pairing mechanism can no longer be classified as spin-singlet or spin-triplet. This "parity mixing" is the  distinguishing feature of  non-centrosymmetric superconductors.

One might expect parity mixing to have a dramatic effect on the macroscopic behavior.  If time-reversal $T$ is also broken (either spontaneously or, for example, by applying an in-plane magnetic field to a 2D film), then the most obvious implication is the "superconducting magnetoelectric coupling": an additional contribution in the London equation for the current that is linear both in  the magnetic field and the superfluid density. Within the Ginzburg-Landau (GL) description, it arises from the Lifshitz term in the free energy  which is of the first order in spatial derivatives of the order parameter \cite{MineevSamokhin,Edelstein}. It has been argued \cite{Agterberg,KaurPRL2005} that this leads to a helical superconducting state (akin to the Fulde-Ferrell state \cite{FF,LO}), where the phase of the order parameter depends linearly on coordinates and thus the Cooper-pair condensate has a non-vanishing momentum \cite{Agterberg,KaurPRL2005}, allowed by broken $P$ and $T$ symmetries. However, an experimental confirmation of the helical state has remained elusive. Recently a Superconducting Diode Effect (SDE) was observed in  Nb/V/Ta superlattices, controlled by an in-plane magnetic field \cite{AndoSDE}. A related Giant Josephson Diode Effect (JDE) was observed in a niobium-semimetal-niobium Josephson junction \cite{JDE}. Both were interpreted in terms of a condensate with a nonzero Cooper-pair momentum induced by the magnetic field. 

Here we will discuss the helical state and its experimental manifestations as well as other effects arising in $P$-  and $T$-violating superconductors. In particular, we will argue on general gauge-invariance grounds that the effects of the leading Lifshitz coupling  are absent in a uniform bulk state, and thus a helical superconductor is not really macroscopically distinct from a conventional superconductor. The effects of the equilibrium Cooper-pair momentum, a clear signature of $P$ and $T$ violation, can 
and do, however appear in certain mesoscopic geometries or non-uniform states, such as interfaces (e.g., JDE), non-simply-connected samples, vortex lattices, and, most strikingly, dislocations.  We also  identify another effect, a strain-induced Josephson current, as a macroscopic manifestation of $P$ and $T$ breaking in a superconducting state.   
\subsection{Manuscript organization and results}

The content of the paper and our results are as follows. In Section II we use the standard model for $P$ and $T$ broken superconductors \cite{Smidman2017} and gauge-invariance to argue that  physical manifestations of the leading spatially uniform Lifshitz coupling and the associated Cooper-pair momentum are hidden in the bulk, only revealed in specialized mesoscopic geometries that we discuss. Building on a GL model, we  propose and analyze a simple generalization that describes the coupling of the superconducting order parameter and the elastic degrees of freedom allowed by broken inversion and time reversal symmetries - piezo-superconductivity. One of the results of this analysis is that a dislocation binds a fractional flux proportional both to the Lifshitz coupling and the Burgers vector. We also explain an analogy between superconductors and insulators based on the presence of approximate higher-form symmetries in both classes of materials. In the light of this observation, the Lifshitz coupling is interpreted as a chemical potential for a higher-form winding symmetry, while the helical state itself is analogous to a pyroelectric insulator. We use this higher-form symmetry formulation to support our conclusion that a uniform helical superconductor is not a bulk state that is distinct from a conventional superconductor. In Section III we predict another manifestation of $P$- and $T$-breaking superconductivity, namely a  current-dependent strain (at zero voltage)  in a Josephson junction and a complementary strain-induced phase difference across the junction.  In Section IV we discuss the effect of elastic coupling on the nature of the normal-superconductor transition, concluding that no new effects arise beyond the "magneto-elastic" coupling studied in magnetic critical systems, that can drive the transition first-order or give well studied "Fisher renormalization" of the critical exponents. We also analyze elastic anomalies associated with piezo-superconductivity around finite temperature phase transition into the superconducting state. We conclude in Section V with an  overview of our predictions and consider prospects for their experimental detection.

\section{Piezo-superconducting free energy}

\subsection{Ginzburg-Landau free energy with Lifshitz coupling}

Within the Ginzburg-Landau phenomenological approach, sufficient for near-equilibrium static phenomena, the low-energy physics of a superconductor is described by a complex scalar $\psi(\rv)=|\psi(\rv)|e^{i\varphi(\rv)}$. Under a gauge transformation which maps the external vector potential $\bf A$ to $\bf A+\nabla\chi$, the Cooper-pair field $\psi$ has charge $2e$, transforming according to $\psi\mapsto\psi e^{2ie\chi}$. The free energy density can be expanded as a power series in a  gauge-invariant (covarient)  derivative  $D_a\psi=(\nabla_a-2ie A_a)\psi$:
\begin{equation}\label{GL}
f=V(\psi^*\psi)-2{\rm Re}\left(i\kappa_{ab} h_a \psi^*\cD_b\psi\right)+{\rm Re}\left(\kappa_{ab} D_a\psi^* \cD_b\psi\right)+\ldots,
\end{equation}
where we included the Lifshitz (second) term characteristic of non-centrosymmetric superconductors in addition to the conventional Ginzburg-Landau form. The ellipsis denote subdominant contributions with higher derivatives and  powers of $\psi$. The Ginzburg-Landau potential $V$ is also invariant under the above $U(1)$ gauge transformation. Although it is usually taken to be a quartic function of $\psi$, its detailed form will not be important here. If one imposes both the spatial inversion symmetry $P$ and the time-reversal symmetry $T$, then  $h_a$ has to vanish, while the tensor $\kappa_{ab}$ must be real and symmetric. 

In this paper we are interested in the situation where either $P$ or $T$ (or both) are broken, and then the vector $\bh$ may be complex while the tensor $\bkappa$ may be complex and Hermitian. However, it is easy to see that the imaginary part of $\bh$ contributes only a total derivative to $f$ and thus can be dropped. Similarly, the imaginary part of $\bkappa$ contributes a total derivative plus a term of the form $2e\, {\rm Im}\,\kappa_{ab} F_{ab} |\psi|^2$, where $F_{ab}=\partial_a A_b-\partial_b A_a$. Such a term corresponds to a spontaneous magnetization and does not concern us here. In what follows we will take the vector $\bh$ to be real, the tensor $\bkappa$ to be real and symmetric, and both are spatially uniform. Stability also requires $\bkappa$ to be positive-definite. 

The vector $\bh$ is a polar vector which is odd under time-reversal. Thus, it can be nonzero only if both $P$ and $T$ are broken. Such a term has apparently been first  proposed by Edelstein \cite{Edelstein}; a similar extension for a two-component order parameter has appeared even earlier \cite{MineevSamokhin}. The physics associated with $\bh$ has been extensively discussed in the context of non-centrosymmetric superconductors in an external magnetic field, where $P$ is broken spontaneously while $T$ is broken explicitly by the external magnetic field \cite{Edelstein,Agterberg,KaurPRL2005}. In particular, it has been predicted in Ref. \onlinecite{Agterberg} that a nonzero $\bh$ leads to a ``helical'' superconducting state, where the phase of $\psi$ depends linearly on coordinates akin to the Fulde-Ferrell\cite{FF,LO,LRreviewFFLO,LRreviewFFLO2} superconductor (though, with distinct symmetries and physical mechanism). Indeed, the free energy can be re-written as 
\begin{equation}\label{GLfreeenergy}
f = V(\psi^*\psi)+\kappa_{ab} |(\cD_a + i h_a)\psi|^2 - \kappa_{ab} h_a h_b |\psi|^2 +\ldots ,
\end{equation}
so minimizing the kinetic energy in the absence of electromagnetic field and with free boundary conditions gives $\psi\sim e^{-i\bh\cdot  {\mathbf r}}$. Thus, $\bh$ is often interpreted \cite{Smidman2017} as the equilibrium momentum of the Cooper-pair condensate. 

We note that, apart from a upward shift in the transition temperature (quadratic in $\psi$ correction to $V$) in Eq. (\ref{GLfreeenergy}), $\bh$ enters only through an "extended  covariant derivative" $\cD_a+ih_a$. This suggests a geometric interpretation of $\bh$ as a constant $U(1)$  vector potential, i.e., a flat background $U(1)$ gauge field. As is well known from the theory of the Aharonov-Bohm effect, only the holonomy of a flat gauge field along a non-contractible loop is physically meaningful. Since the physical space $\RR^3$ has no non-contractible loops, this  implies that in the bulk $\bh$ can be ``gauged away''. Indeed, a constant $\bh$ can be absorbed into a redefinition $\psi\mapsto \psi e^{-i \bh\cdot {\mathbf r}}$ (crucially, at the expense of modifying terms with higher derivatives, the leading one quadratic in $\psi$, with three derivatives). Importantly, the potential  $V(\psi^*\psi)$ is unchanged thanks to ordinary electromagnetic gauge-invariance. More generally, a non-constant $\bh$ which is a gradient of a function can be removed by a redefinition of $\psi$, and our conclusions also extend to this case.  Conversely,  $\bh$ with a nonzero curl will have physically observable effects. 

This conclusion does not imply that a constant nonzero $\bh$ has no observable effects. Indeed, there are a number of mesoscopic experimental configurations where  $\bh$ has  physical consequences. One is the  familiar Little-Parks experiment, where a sample has a non-simply-connected shape \cite{DimitrovaFeigelman}, e.g., a ring or an annular Corbino geometry, so that $\bh$ cannot be ``gauged away''. A nonzero $\bh$ will shift the Little-Parks oscillations of the current as a function of the magnetic flux by the holonomy of $\bh$, which is a clear sign of $P$ and $T$ violation. Another example is that discussed in Ref. \onlinecite{KaurPRL2005}, where the helical ground state affects the Josephson current between a superconductor with a nonzero $\bh$ and an ordinary superconductor with $\bh=0$. It leads to oscillations of the DC Josephson current as a function of the extent of the junction in the direction of $\bh$, suppressing the overall current. This physical effect is consistent with the  above interpretation of $\bh$ as a gauge field, since $\bh$ along the plane of the interface, with a step across it, has a non-vanishing curl and therefore cannot be "gauged away". The aforementioned Josephson Diode effect \cite{JDE} is another example of a mesoscopic boundary effect, consistent with 
above discussion.
In all these examples, the dependence on $\bh$ is periodic, with period $2\pi/L$, set by the mesoscopic geometry of the structure.
This is consistent with our above observation that $\bh$ can be gauged away 
from the macroscopic bulk modulo $2\pi/L$, entering only through  boundary conditions of a finite size sample. We also note that in the London limit (constant $|\psi|$), the Lifshitz coupling is a total derivative, $2\kappa_{ab} h_a\cdot\nabla_b\varphi$, where $\varphi={\rm arg}\,\psi$.

Since, as argued above, the leading Lifshitz coupling has negligible effects in the thermodynamic limit, 
the bulk physics of a superconductor with broken $P$ and $T$  symmetries must be controlled by higher order terms, that might have otherwise been viewed as subdominant. 
At quadratic order in the Cooper-pair order parameter $\psi$, the leading such term is cubic in covariant derivative, given by 
\begin{equation}
f_{23}=
2{\rm Re} \left( i \gamma_{abc} \psi^*\cD_a\cD_b\cD_c\psi \right),
\label{f23}
\end{equation}
where the real tensor $\gamma_{abc}$ is odd under $P$ and $T$ symmetries. Without loss of generality we take the tensor  $\gamma_{abc}$ to be completely symmetric. Indeed, since $[D_a,D_b]\psi=-2ie F_{ab}\psi$, the components which are not completely symmetric can be absorbed into a $\bB$-dependent redefinition of $h_a$. 

Another $P$ and $T$-violating leading higher order term, quartic in $\psi$ and linear in covariant derivative, is
\begin{equation}
f_{41}= -i  \kappa_{ab} h'_a|\psi|^2\left(\psi^* D_b\psi-\psi D_b\psi^*\right). 
\label{f41}
\end{equation}
Including such a term in the free energy is equivalent to allowing $\bh$ to depend on $|\psi|^2$, and for uniform $|\psi|$ and $h'$ can be gauged away in the bulk per our above discussion of shifting away the Lifshitz term.
$f_{23}$ and $f_{41}$ are   irrelevant at the transition into the superconducting state, but lead to vortex anisotropy, akin to that  recently observed in Ref. \onlinecite{anisotropy}.

In the  following sections we will describe another striking manifestation of $\bh$. Namely, we will show that in the presence of a dislocation, the physics is sensitive to  $\bh$ modulo the reciprocal lattice vectors (much larger than the mesoscopic $2\pi/L$ manifestations discussed above).  We will also show that including the interaction between $\psi$ and the elastic degrees of freedom gives rise to further $P$ and $T$-violating effects in a non-centrosymmetric superconductor.

\subsection{Coupling to elastic degrees of freedom}
When looking for other possible effects of $P$ and $T$ breaking on superconductivity, we note that boundaries are not the only kind of inhomogeneities that can occur in a crystal. Another ubiquitous type of inhomogeneity is a dislocation  - a translational topological defect in a crystal's lattice structure. To understand how dislocations interact with superconductivity, we need to take into account the coupling of $\psi$ to the elastic degrees of freedom. To do this, it is useful to regard phonons as Goldstone bosons of the spontaneously broken translation symmetry. The guiding principle of  the free energy construction is its invariance  under arbitrary rotations and translations of the physical space, as well as under the crystallographic space group in reference space. This is accomplished by introducing Goldstone fields $X^i(\rv)$, $i=1,2,3$ which are functions of the "material  coordinates" $r^a$, $a=1,2,3$. In equilibrium we have $X^i=O^i_a r^a+X_0^i$, where $O^i_a$ is an orthogonal matrix and ${\bf X}_0$ is an arbitrary vector. More generally, one regards the functions $X^i(\rv)$ as describing a map from the three-dimensional ``reference space'', parameterized by the material coordinates $r^a$ to the three-dimensional physical "target" space $X^i$. This map is one-to-one if crystal defects (dislocations and disclinations) are absent, otherwise it is {\it locally} one-to-one.

Note that the indices $a,b,c,...$ label the coordinates of the reference space, while the indices $i,j,k,...$ label the coordinates of the physical space. Elements of the crystal point group act on the former, while rotations of the physical space act on the latter. The "elastic vielbein" - the Cauchy deformation
tensor $E^i_a=\frac{\partial X^i}{\partial r^a}$ and its inverse $\tE^a_i=\frac{\partial r^a}{\partial X^i}$ allow one to convert one type of index to the other. It is useful to introduce a reference-space metric tensor $g_{ab}=E^i_a E^i_b$, where summation over repeated indices is understood. We note that  $g_{ab}$ is a tensor in the reference space and a scalar in the physical space. Geometrically, it is a pull-back of the flat metric on the physical space to the reference space via the map $\br\mapsto\bX(\br)$. In equilibrium,  $g_{ab}=\delta_{ab}$ is flat, so the conventional "right
Cauchy-Green" strain tensor, which is invariant with respect to rotations in the physical space,  $u_{ab}=\frac{1}{2}(g_{ab}-\delta_{ab})$ vanishes in an undeformed crystal. 

In the ``Lagrangian" approach, the field $\psi$ is regarded a function of the reference coordinates $r^a$ (and time, if one is interested in time-dependent problems). The free energy density expanded to quadratic order in derivatives of $\psi$ now takes the form
\begin{equation}
f=V(\psi^*\psi)-i\kappa_{ab} h_a[{\bf u}] \left(\psi^*\cD_b\psi-\psi\cD_b\psi^*\right)+\kappa_{ab} \cD_a\psi^* \cD_b\psi+ f_{\rm el} +\ldots,
\label{elasticCoupling}
\end{equation}
where $D_a\psi=\partial_a\psi-2e i \partial_a X^i A_i\psi$ and
$f_{\rm el} = \frac{1}{2} c_{ab,cd}u_{ab}u_{cd}$ is the usual elastic free energy density, which is a function of the symmetric strain only, and ellipsis stand for higher order terms that are subdominant for small strains and currents. Note that the vielbein appears in the coupling to the vector potential to ensure rotational invariance in the physical space. In the Lagrangian description, $\partial_a\varphi$ and $A_i$ are vectors in a reference and a physical space, respectively. In both the Eulerian and Lagrangian descriptions the velbein in the minimal coupling ensures that the gauge transformation takes a standard form, $\psi\rightarrow \psi e^{2ie\chi}$, $A_i \rightarrow A_i + \partial_i\chi$. 

The symmetric tensor $\kappa_{ab}$ and the vector $h_a$ may themselves be functions of the strain tensor $\bf u$. For example, the dependence of $\kappa_{ab}$ on $\bf u$ accounts for a possible dependence of the London penetration length on the strain. For our purposes here, it will be sufficient to regard $\kappa_{ab}$ as strain-independent and expand $h_a[{\bf u}]$  to linear order in the strain, $h_a[{\bf u}]=h_a + g_{b,cd}u_{cd}$.

After some rearrangement and a negative correction to the elastic energy $\delta f_{\rm el} = - \kappa_{ab}(g_{a,cd} u_{cd} + h_a)(g_{b,ef}u_{ef}+h_b)|\psi|^2$ encoding a superconductivity-induced elastic distortion and strain-induced shift in the superconducting transition temperature $T_c$, we get
\begin{equation}
f=V(\psi^*\psi)+\kappa_{ab} \left(\cD_a\psi+i (h_a + g_{a,cd}u_{cd})\psi\right)^* \left(\cD_b\psi+i(h_b+g_{b,ef}u_{ef})\psi\right) + f_{\rm el} + \delta f_{\rm el}+ \ldots .
\label{elasticCoupling2}
\end{equation}
At first sight, it might seem that a constant vector $\bh$ can again be absorbed into $\psi$ by redefining $\psi\ra\psi e^{-i \bh\cdot {\mathbf r}}$. However, as discussed in the Introduction, this is not true if there are boundaries or if the sample is not simply-connected. We will now show that it is also not true in the presence of a dislocation.

\subsection{Dislocation probe of the Lifshitz coupling}

Usually, dislocations are incorporated by allowing the map $\bX\mapsto \br(\bX)$ from the physical space to the reference space to be multi-valued, so that different ``branches'' of the map are related by shifts in the Bravais lattice of the crystal. As a result, loops in the physical space encircling dislocations map to non-closed paths in the reference space whose endpoints are related by a shift in a  reference space lattice - the Burgers vector {\bf b} of the dislocation. The inverse map $\br\mapsto \bX (\br)$ is single-valued, but has the property that points in the reference space separated by the Burgers vector are mapped to the same point in the physical space.

Since $\psi$ must be well-defined on the physical space, the redefinition $\psi\ra\psi e^{-i \bh\cdot {\mathbf r}}$ is not "harmless"
in the presence of dislocations. Namely, in a dislocated sample a nonzero $\bh$ (a Lifshitz term) in the free energy (\ref{GLfreeenergy}),(\ref{elasticCoupling2}) cannot be shifted away without consequences, and therefore has physical effects in the bulk, probed by a dislocation. To see what these effects are, we may switch to the Eulerian description and express all the fields as functions on the physical space. After defining $\psi=\tilde\psi e^{-i \bh\cdot {\mathbf r}}$, the winding number of ${\rm arg}\,\tilde\psi=\tilde\varphi$ around the dislocation will no longer be integral. Instead, for a closed curve $\gamma$ linking the dislocation line we will have
\begin{equation}\label{winding}
\oint_\gamma  \nabla\tilde\varphi\cdot d\ell=\bh\cdot \bb+2\pi n,
\end{equation}
where $\bb$ is the Burgers vector of the dislocation and $n$ is an arbitrary integer. 

One physical consequence of the shift (\ref{winding}) of the winding number of $\tilde\varphi$ is that when a $T$ and $P$-breaking material is cooled through a superconducting phase transition, a dislocation with a Burgers vector $\bb$ will necessarily trap a fractional magnetic flux of magnitude $(\bh\cdot \bb)/2e$. 
This prediction should be testable via scanning tunneling and Hall probe microscopies.

From a geometric viewpoint, it may be more natural to  identify points in the reference space related by Bravais lattice vectors so that the map $\bX\mapsto \br(\bX)$ is  uni-valued and maps loops to loops. It still has singularities at the locations of dislocations, so that the complement of the singularity locus is topologically non-trivial. In such description the reference space  becomes topologically non-trivial (has non-contractible loops). The vector $\bh$ can be interpreted as a flat $U(1)$ gauge field on the reference space, but since the space is no longer simply-connected, its holonomy (i.e. Wilson loops) is now observable.  Homology classes of non-contractible loops on such reference space are then labeled by elements of the Bravais lattice. The holonomy of the gauge field $\bh$ along a loop labeled by a Bravais lattice vector $\bb$ is therefore given by $e^{-i \bh\cdot\bb}$.  We note, however, that although $\bh$ is thereby probed by dislocations, its values that differ by an element of the reciprocal lattice are physically indistinguishable, i.e., periodically identified by reciprocal lattice vectors.

Identification of values of $\bh$ related by a vector in the reciprocal lattice suggests that $\bh$ should be interpreted as the net {\em quasi}-momentum of a Cooper pair, consistent with the {\em discrete} translational symmetry of 
a crystal.
Finally we note that this distinction between momentum and quasi-momentum is important only because of discreteness on atomic scale, which is why a consideration of dislocations is needed to identify  $\bh$ as the {\em quasi}-momentum.

In closing we observe a close similarity between a dislocation trapping a fractional flux and a mesoscopic superconducting ring in a Little-Parks experiment. Although both are associated with an Aharonov-Bohm phase, in the case of a dislocation there is no actual hole in the physical (sample) space, and the nontrivial topology is associated with the holonomy in the reference space. To the extent that vortices carrying a flux quantum can enter and leave the system, in equilibrium a superconductor is  sensitive only to a fraction of a flux quantum set by $\bh$ modulo the reciprocal lattice. However, because equilibration of persistent currents can be slow on experimental time scales, multiple flux quanta can be induced on a dislocation by a fast ramp of $\bf h$ (via e.g., an in-plane magnetic field) beyond its fractional value. 

\subsection{Higher-form symmetry characterization of helical superconductors and pyroelectric insulators}

In this section we propose a higher-form symmetry-based characterization of  helical superconductors. The key fact is that $\varphi$ is a periodic scalar, $\varphi\sim\varphi+2\pi$. Therefore for any non-trivial degree-1 homology class $\gamma$ there is an integral conserved quantity, the winding (vortex) number  of $\varphi$ defined as $\frac{1}{2\pi} \int_\gamma d\varphi$. Since this conserved quantity is obtained by integrating over a closed curve - a codimension $d-1$ submanifold of the  $d$-dimensional spatial slice - the corresponding symmetry is a $(d-1)$-form symmetry \cite{higherform}.

Conservation of the winding number can be thought of as a local conservation law if we  introduce a 1-form $J=d\varphi$ on the $(d+1)$-dimensional space-time which (in the absence of vortices, namely for an exact $1$-form  $J$) tautologically satisfies $dJ=0$. Equivalently, if we define the Hodge dual current $\tilde J = *d\phi$ (whose components are given by $\tilde J_{\mu_1,\ldots \mu_d} =\epsilon_{\mu_1,\mu_2,\ldots \mu_{d+1}}\partial_{\mu_{d+1}}\varphi$), then it is divergenceless, $\partial_{\mu_1} {\tilde J}^{\mu_1\ldots\mu_d}=0$. Letting $d=3$ for definiteness, we can write $dJ=0$ in a suggestive form by defining a vector field $\bJ$ with components $J_i=\partial_i\varphi$ and the chemical potential $\mu=J_0=\partial_0\varphi$. Then $dJ=0$ is equivalent to
\begin{equation}\label{localJ}
    \nabla\times{\bf J}=0, \quad
    \partial_t{\bf J} = \nabla\mu.
\end{equation}
The second equation implies that for any closed curve $\gamma$ the integral $\oint_\gamma \bJ\cdot d{\mathbf \ell}$ is independent of time. 
Its physics is a local statement of the Josephson equation.
For a finite line-segment, this equation equates the  increase with time of phase winding across the line-segment with the difference of chemical potentials between the two ends. The first equation in (\ref{localJ})  implies that $\oint_\gamma \bJ\cdot d{\mathbf \ell}$  does not change under deformations of $\gamma$, which is part of the definition of the ($d-1$)-form symmetry. In particular, it can be nonzero only if $\gamma$ cannot be shrunk to a point. 

We note that this $(d-1)$-form symmetry becomes exact only in the London limit, where $|\psi|$ cannot vanish. If vortices are allowed, then $\varphi={\rm arg}\, \psi$ is ill-defined at the vortex cores where $\psi=0$, and $\nabla\times \bJ =n_v$ is given by the vortex density. If vortices can be regarded as non-dynamical, one can restore the $(d-1)$-form symmetry by regarding all fields as defined outside the vortex cores. This is a reasonable definition in the London limit of vortices frozen inside a sample by e.g., pinning defects.

The 1-form $J$ can be naturally coupled to an external $d$-form gauge field $a=\frac{1}{d!} a_{\mu_1\ldots\mu_d} dx^{\mu_1}\ldots dx^{\mu_d}$ by adding a term $\int a\wedge J=\frac{1}{d!}\int a_{\mu_1\ldots\mu_d} {\tilde J}^{\mu_1\ldots\mu_d} d^{d+1} x$ to the action. Since $dJ=0$, this coupling is invariant under a gauge symmetry $a\mapsto a+d\lambda,$  where $\lambda$ is a $(d-1)$-form. In particular, an analog of the chemical potential for a $(d-1)$-form symmetry is an external gauge field whose purely spatial components vanish while the the components $a_{0j_1\ldots j_{d-1}}$ are constant. Taking $d=3$ for definiteness, this corresponds to shifting the action by 
$\frac{1}{6}\int d^3x\, \eps_{ijk} a_{0jk}\partial_i\varphi$ for a constant skew-symmetric tensor $a_{0jk}=\eps_{ijk} a_i$. Comparing with Eq. (\ref{GL}), we observe that in the London limit where $|\psi|^2$ is a constant the vector $\bf a$ is related to $\bh$ of the Lifshitz coupling:
\begin{equation}
a_i=
6\partial_a X^i\kappa_{ab} h_b |\psi|^2.
\end{equation}
Thus in the London limit $\bh$ is identified with the chemical potential for the winding number.


The concept of a chemical potential for a higher-form symmetry may seem unfamiliar, but in fact the physics of insulators can be interpreted in similar terms. Recall that an insulator is characterized by the ability to define vector fields $\bD$ and $\bH$ which in the absence of external sources satisfy Maxwell's equations for a dielectric,
\begin{equation}\label{maxwell}
\nabla\cdot\bD=0,\quad \frac{\partial\bD}{\partial t}= \nabla\times\bH.
\end{equation}
Similarly to eq. (\ref{localJ}), the first equation ensures that for a closed  $d-1$-dimensional surface $\Sigma$ the quantity $\int_\Sigma \bD\cdot d\bS=0$ is independent of the choice of $\Sigma$ within its homology class. The second equation is the  continuity equation for the displacement field $\bD$. This means that an insulator possesses a 1-form symmetry (for all $d$).  This 1-form symmetry is typically only approximate, since  physical systems have a non-vanishing conductivity at positive temperatures. It becomes exact only in the limit when the energy or mobility gap for charged excitations is infinitely large, or when the temperature goes to zero.

Just as the Lifshitz coupling  $\int d^3x\, \bh\cdot \bJ$ in a helical superconductor free energy modifies the chemical potential for the winding number, adding a term $\int d^3x\, \bq\cdot \bD$ to the dielectric free energy (where $\bq$ is a constant vector) changes the chemical potential associated with the dielectric 1-form symmetry. Such a term shifts the equilibrium value of the electric field in the insulator by $\bq$ and gives rise to a spontaneous electric polarization. This is known as a pyroelectric state. In the case of a helical superconductor, adding a chemical potential for the winding number results in a nonzero equilibrium value of the superfluid current $\bJ$. Thus a helical superconductor is to an ordinary superconductor as a pyroelectric is to an ordinary insulator. 


We note that an equilibrium current density in a helical superconductor is not in conflict with the Bloch-Bohm  theorem \cite{Bloch}. This theorem states that the net equilibrium current through a section of a quasi-1d system vanishes when its length is taken to  infinity while the width is kept finite. The equilibrium current density in a helical superconductor  is given by the minimal norm of the vector field $\cD_a\varphi+h_a$. Since $\varphi$ is allowed to have a nonzero winding number, in a quasi-1d system of length $L$ and a cross-section area $A$ the minimizing configuration has the net current which scales like $A/L$. This vanishes in the limit $L\ra\infty$, in agreement with the Bloch-Bohm theorem.

A further similarity between pyroelectrics and helical superconductors is that neither represents a macroscopic phase of matter which is truly distinct from an ordinary superconductor or an ordinary insulator, respectively. In the case of pyroelectrics, a spontaneously generated electric polarization is almost canceled in the  long-time limit by a flow of free charges arising from a non-vanishing conductivity. Screening is not perfect only because the free charge is quantized, while the spontaneous electric polarization and the associated surface charge, in general, are not. However, the residual electric polarization is microscopic, rather than macroscopic, with polarization density vanishing in the thermodynamic limit. In the case of helical superconductors, an analogous process takes place when one starts out with a non-vanishing persistent current density: thanks to vortex-anti-vortex nucleation, the winding number of $\varphi$ changes until the current $\cD_a\varphi+h_a$ reaches the minimum allowed by the sample topology. The residual current is microscopic (more precisely, of order $A/L$, as explained in the previous paragraph). As a result, all effects of $\bh$ on bulk properties are periodic with period of order $1/L$ and thus vanish in the thermodynamic limit. 
Ultimately, the reason for this is that, strictly speaking, neither the 1-form symmetry of insulators nor the $(d-1)$-form symmetry of superconductors are exact. The former is violated by a nonzero charge conductivity, while the latter is violated by nonzero vortex  mobility. 

\subsection{T-invariant non-centrosymmetric superconductors in a magnetic field}

If a external magnetic field ${\bf B}$ {\it uniformly}  penetrates a superconductor (realized in a film subjected to an in-plane magnetic field), then uniform $h_a$ and $g_{a,bc}$ are induced  through the explicit $T$-breaking. Spatial inversion $P$ must still be broken, but this occurs naturally if the two surfaces of the film are not equivalent. To linear order in $\bB$ we must have
\begin{equation}
h_a=\alpha^b_a B^i\partial_b X^i,\quad g_{a,bc}=\beta^d_{a,bc} B^i \partial_d X^i.
\end{equation}
Here $\alpha^b_a$ is a $T$-even and $P$-odd rank-2 tensor, and $\beta^d_{a,bc}$ is a $T$-even and $P$-odd rank-4 tensor which is symmetric in the reference-space indices $b,c$. Above we took into account that ${\bf B}$ is a vector in the physical space, and so used the elastic vielbein to ensure rotational invariance in the physical space.  The form of the tensors $\alpha^b_a$ and $\beta^d_{a,bc}$ is constrained by the crystal symmetry. For example, for the $C_{4v}$ point group one must have $\alpha^b_a\sim \eps_{ab}$, for the plane of the film transverse to the 4-fold axis.

In the presence of ${\bf B}$ the Ginzburg-Landau free energy depends not only on the strain $u_{ab}$ but also on the vielbein $E_a^i=\partial_a X^i$. Physically, this encodes the dependence of the free energy on the orientation of the superconducting crystal with respect to the in-plane magnetic field. Ordinary magnetization energy $-{\bf M}\cdot {\bf B}$ also contributes to such a dependence, however. 

We note, however, that in a case of a nonuniform magnetic field ${\bf B}(\rv)$,  Lifshitz coupling $\bh(\rv)$ may have a nonzero curl and then cannot be gauged away. An important example of this is a vortex state of a Type II superconductor. Such a non-uniform Lifshitz coupling may therefore lead to anisotropy and a field-driven distortion of the vortex lattice \cite{anisotropy,yerin}.



\section{Elastic Josephson effects}

As discussed in previous section, in a simply-connected sample and in the absence of dislocations, the effect of the vector $\bh$ can be eliminated by redefining $\varphi$. The rank-3 tensor $g_{a,bc}$, on the other hand, cannot be eliminated. It is odd under both $P$ and $T$. In other words, its symmetry properties are exactly the same as those of the conventional piezomagnetic coupling. When it is non-zero, it results in an additional contribution to the elastic stress proportional to the superfluid velocity $v_a=D_a\varphi=\partial_a\varphi-2e \partial_a X^i A_i$:
\begin{equation}
\sigma_{cd}=\frac{\partial f}{\partial u_{cd}}=c_{ab,cd} u_{cd}-2\kappa_{ab} g_{b,cd} |\psi|^2 D_a\varphi .
\end{equation}
Reciprocally, the London equation for the superconducting current now has a contribution from the strain:
\begin{equation}\label{supercurrent}
j_i=4e\kappa_{ab} |\psi|^2 E^i_a (D_b\varphi-g_{b,cd}u_{cd})\simeq 4e\kappa_{ij}|\psi|^2 D_j\varphi-4e\kappa_{ib}|\psi|^2 g_{b,cd} u_{cd}.
\end{equation}
The first term in this equation is the usual London current proportional to the superfluid "density" tensor $\kappa_{ij}$ with respect to the ``physical'' coordinate axes. The second term reflects a novel effect of a generation of a dissipationless current by strain. It  leads to the "elastic" Josephson effect as we now discuss.

Consider a Josephson weak link arising from a narrow constriction ("bridge") in a superconducting material. 
The usual Josephson current has the form $I=I_c \sin\Delta\varphi$, where $\Delta\varphi$ is the change of the phase $\varphi={\rm arg}\,\psi$ across the "bridge". 
Suppose a strain is induced in the bridge by an external stress. This changes the value of $\bh[u]$ and therefore the net momentum of the Cooper pairs. For an approximately uniform strain, the  additional phase accumulated by a Cooper pair as it traverses the bridge is $L g_{z,ab}u_{ab}$, where $z$ is the axis along the bridge. More generally, this is an order-of-magnitude WKB-like estimate, since the actual strain depends on coordinates and to determine the numerical factor requires a solution of the full elasticity problem. For concreteness, let us consider the case where the bridge is twisted by an angle $\theta$ by an external torque $\tau$. If $L\gg W$, then the components $u_{zx}$ and $u_{zy}$ of the strain are of order $\theta W/L$ while the other components are negligible. The Josephson current then takes the form
\begin{equation}
I=I_c \sin(\Delta\varphi+ k\theta),
\end{equation}
where $k\sim W g_{z,za}$, and $a=x,y$.
This current corresponds to the following energy function which depends on both $\Delta\varphi$ and $\theta$:
\begin{equation}
V(\Delta\varphi,\theta)=-\frac{1}{2e} I_c\cos\left(\Delta\varphi + k\theta\right) +\frac{c\theta^2}{2L} - \tau\theta,
\end{equation}
Here $c$ is the torsional rigidity of the bridge given by a geometric factor times the shear modulus times the square of the cross-sectional area, defined to make the dependence of the elastic energy on $L$ explicit. The Josephson current is given, as usual, by $I=2e\frac{\partial V}{\partial\Delta\varphi}$.

In the above discussion $\Delta\varphi$ was taken as the independent control parameter. However, from the experimental viewpoint, it is more natural to fix and tune the external current $I$ through device. The corresponding potential is the Legendre transform of $V(\Delta\varphi,\theta)$ from $\Delta\varphi$ to $I$, and is obtained by adding a term $-\frac{I\Delta\varphi}{2e}$ to the potential $V(\Delta\varphi,\theta)$ above and extremizing it with respect to  $\Delta\varphi$ and $\theta$. This gives 
\begin{eqnarray}\label{theta}
\theta&=&\frac{L}{c}\tau - \frac{kL}{2e c} I,\\ \label{varphi} \Delta\varphi&=&\Delta\varphi_0 -\frac{kL}{c} \tau + \frac{k^2L}{2e c} I,
\end{eqnarray}
where $\Delta\varphi_0$ is defined by $I=I_c\sin\Delta\varphi_0$.

According to Eq. (\ref{theta}), the coupling $g_{a,bc}$ results in a current-dependent strain as measured by the twist angle $\theta$. Importantly, his behavior is not masked by piezolectric effects since there is no voltage drop across the junction in the DC regime. Furthermore, other effects such as the dependence of the superfluid stiffness $\kappa$ on the strain cannot lead to a linear dependence of $\theta$ on the current $I$. Such a linear dependence is a smoking gun for $T$-odd effects. Conversely, Eq. (\ref{varphi}) shows that $\Delta\varphi$ is affected by the applied stress, the torque $\tau$. While $\Delta\varphi$ is not directly observable, its time-derivative is proportional to the voltage drop and is observable. The AC voltage drop resulting from an AC torque (stress) applied to the junction is distinguishable from the piezoelectric voltage by its linear dependence on the frequency (this, again, is a robust consequence of the fact that $g_{a,bc}$ is $T$-odd).

While the above discussion was phrased for superconductor-constriction-superconductor  junctions, qualitatively the same effects will take place in SIS and SNS junctions provided there is both a $T$ and $P$-breaking in the junction. In fact, it might be easier to observe elastic Josephson effects in such junctions because then the superconducting leads may be taken as ordinary $P$ and $T$-symmetric superconductors. For example, following Ref. \onlinecite{JDE}, one can use a weak link involving a Dirac semi-metal. Spin-orbit interaction will then ensure  the $P$-breaking in the junction, while an in-plane magnetic field will induce the requisite $T$-breaking. 

\section{Normal-to-superconducting transition}

Having  studied the piezo-superconducting state, we now turn to the transition from a metal to a piezo-superconductor. We recall that a conventional thermal 3d  metal-superconductor transition (NS), when continuous, is believed to be in the inverted XY universality class \cite{DasguptaHalperin,HalperinLubenskyMa,LReurophysLett}. The distinguishing $P$ and $T$ violating bulk terms, (\ref{f23}), (\ref{f41}) are irrelevant at the NS critical point, and thus the universality class of the NS  transition would remain unchanged, were it not for the elastic coupling that we explore below. 

\subsection{The effect of the elastic modes on the superconducting transition}

We consider the effect of phonon coupling in (\ref{elasticCoupling}) on the phase transition into a  piezo-superconducting state. Fluctuations of the  superconducting order parameter $\psi$ are described by the standard GL free-energy density $f_{GL} = \kappa|(-i\nabla - 2e {\bf A})\psi|^2 + \alpha|\psi|^2 +\frac{1}{2}\beta|\psi|^4 + \frac{1}{8\pi}(\nabla\times {\bf A})^2$, (neglecting inessential anisotropy in $\kappa$ and for compactness of notation defining $A_a = E^i_a A_i$), supplemented by the Lifshitz coupling
\begin{equation}
    f_{piezo} = -i\kappa (h_a + g_{a,cd}u_{cd})\psi^*(\partial_a - 2e i  A_a)\psi + c.c. ,
\end{equation}
the elastic free-energy density $f_{el}=\frac{1}{2} c_{ab,cd}u_{ab}u_{cd}$, and the strain-density coupling $f_{sd}=\rho_{ab} u_{ab} |\psi|^2$, where $\rho_{ab}$ is a real symmetric tensor.
As discussed in Refs.~\onlinecite{Smidman2017,KaurPRL2005}, for nonzero $h_a$ (arising from spontaneous or explicit $T$-breaking) the transition is to a helical superconductor state with $\psi = \tilde\psi e^{-i {\bf h}\cdot {\bf r}}$, taking place at an elevated critical  temperature, determined by the suppressed coupling $\tilde\alpha =\alpha - \frac{3}{2}\kappa h^2 = 0$, and governed by an effective GL model
\begin{eqnarray}
f &=& \kappa|D_a\tilde\psi|^2 + \tilde\alpha|\tilde\psi|^2 +\frac{1}{2}\beta|\tilde\psi|^4 + \frac{1}{8\pi}(\nabla\times {\bf A})^2+
\frac{1}{2} c_{ab,cd}u_{ab}u_{cd}\nonumber\\
&+& \kappa g_{a,bc}u_{bc}\left[-i\tilde\psi^*D_a\tilde\psi + c.c.\right] 
+ \tilde\rho_{ab}u_{ab}|\tilde\psi|^2 .
\label{ftilde} 
\end{eqnarray}
Here  $\tilde\rho_{ab}=\rho_{ab}-\kappa h_c g_{c,ab}$.
We thus find that, aside from the $T_c$ shift, the standard GL free energy is perturbed by strain-current and strain-density couplings (the last two terms)
\begin{equation}
    f_{u\psi} = (g_{c,ab} j_c + \tilde\rho_{ab} n) u_{ab},
    \label{ujn}
\end{equation}
where $n=|\tilde\psi|^2$ and $j_a=-i\kappa\tilde\psi^* D_a\tilde\psi+c.c$ is the standard number current. 
The electric current is given by 
\begin{equation}
    J_a = 2e\kappa [-i\tilde\psi^*D_a\psi + c.c.] + 4e \kappa g_{a,bc} u_{bc} |\tilde\psi|^2,
    \label{current}
\end{equation}
and contains a strain-dependent contribution. 

It is natural to ask whether these elastic couplings 
modify the critical nature of the metal-superconductor phase transition. At the level of power-counting one can formally integrate out the phonons entering through the strain. This induces a current-current, current-density and density-density couplings,
\begin{equation}
    f_{n,j} = -\frac{1}{2}(g_{e,ab} j_e + \tilde\rho_{ab}n)\hat K_{ab,cd}(g_{f,cd} j_f + \tilde \rho_{cd} n)
    \label{longrange_nj}
\end{equation}
where $\hat K_{ab,cd}$ is a long-range kernel of dimension zero arising from elasticity and  given by the Fourier transform of  $K_{ab,cd}(q)=q_a [q_e c_{eb,fd}q_f]^{-1}q_c$.  Now it is straightforward to see that current-density and current-current couplings in (\ref{longrange_nj}) are irrelevant relative to the short-range ($\beta$ in the standard GL model) and long-range density-density quartic couplings, as they respectively contain one and two extra derivatives. 
The strain-density interaction in (\ref{ujn}) and the corresponding induced long-range density-density coupling in (\ref{longrange_nj}) have been explored in a variety of compressible systems, e.g., supersolids and compressible magnets \cite{BergmanHalperin}. It is known that the nature of the transition depends on the experimental ensemble: a first-order transition at fixed pressure and a continuous transition at fixed volume with  "Fisher-renormalized" \cite{FisherRenormalized} exponents. We thus expect the same effects at the NS transition when the elastic degrees of freedom are taken into account. Since $\rho_{ab}$ is nonzero, in general, these  effects are not special to non-centrosymmetric superconductors.

\subsection{Elastic anomalies}
Conversely, one may consider the feedback of the superconducting phase transition on the phonons. Within mean-field theory (mft) approximation, treating $\tilde\psi = \tilde\psi_0$ as a constant, and integrating out the gauge field, or equivalently  expressing it in terms of the strain using the corresponding Euler-Lagrange equation, $2e A_a = g_{a,bc} u_{bc}$ we find 
\begin{equation}
 \tilde f_{el} = \frac{1}{2} \tilde c_{ab,cd}u_{ab}u_{cd}    - \tilde\sigma_{ab} u_{ab},
\end{equation}
where the effective elastic tensor and superconductivity induced stress are 
\begin{eqnarray}
\tilde c_{ab,cd} = c_{ab,cd} + d_{ab,cd}|\tilde\psi_0|^2,\quad
\tilde\sigma_{ab}&=& -\tilde\rho_{ab}|\tilde\psi_0|^2. \end{eqnarray}
The tensor $d_{ab,cd}$ is independent from the ones previously introduced and  encodes how the elastic constant depends on the superfluid density.
Thus, within mft we predict a distortion, $\delta u_{ab}$, and a change of the elastic constants $\delta c_{ab,cd}$ that sets in non-analytically as $\sim|T-T_c|$ below the superconducting transition.  Beyond the mft approximation these critical corrections $\delta c_{ab,cd}, \delta u_{ab}$ scale with energy density according to $\sim |T-T_c|^{1-\alpha}$, where $\alpha$ is the heat-capacity critical exponent. Again, this behavior is a generic consequence of the coupling the superconducting order parameter $\psi$ to the elastic modes and is not special to non-centrosymmetric superconductors.


\section{Summary and conclusions}

In this paper we have studied superconductors with broken $P$ and $T$ symmetries,
exploring novel couplings between the Cooper-pair condensate and the elastic degrees of freedom that are  special to such materials.
Prior discussions of $P$ and $T$-breaking superconductors highlighted the presence of the single-derivative Lifshitz coupling in the free energy and the properties of the corresponding "helical state". We have argued on very general grounds that, although this coupling $\bf h$ affects certain mesoscopic properties, gauge-invariance precludes its manifestations in a uniform state.  
This can also be understood using an approximate higher-form winding symmetry generic to all superconductors. 

On the other hand, using our generalized Ginzburg-Landau theory, we have argued that the Lifshitz coupling $\bh$ has an interesting effect on dislocations, turning them into vortices carrying a fractional flux. Flux quantization leads to a periodic identification of ${\bf h}$ by  reciprocal lattice vectors. This is in agreement with the interpretation of $\bh$ as the quasi-momentum of the Cooper-pair condensate. We also showed that a novel elastic coupling, special to $P$ and $T$-broken superconductors gives rise to a new effect: a generation of a non-dissipative current by strain. As discussed above, this leads to current-induced strain in $P$ and $T$-breaking Josephson junctions. Finally, we discussed the effects of the elastic modes on the normal-to-superconductor phase transition, and the complementary effects of the phase transition on the elastic modes.

We conclude this paper by estimating the size of some of the effects we described. The authors of Ref. \onlinecite{JDE} interpret their results in terms of a helical state with a quasi-momentum $\bh$ of order of a few inverse-microns. Such a small $\bh$ will give dislocations a magnetic flux whose fractional part (in units of the superconducting flux quantum) is no larger than $10^{-3}$. This is consistent with the weak-link's length being on the order of $10^{3}$ of the elementary dislocation Burgers vector (a lattice constant).  To get up to a higher fraction of the flux quantum, magnetic fields larger than the tiny $12$ milli-Tesla used in these experiments will be necessary, or 
 alternatively, one could look at flux predicted to be necessarily trapped on dislocations in ferromagnetic superconductors such as ${\rm URhGe}$ and ${\rm UCoGe}$ \cite{ferromagneticsuperconductors}.

Turning to the off-diagonal Josephson effect, the elastic strain $u$ induced by a current in a "bridge" is of order
\begin{equation}
u\sim 10^{-6}\left[\frac{g_{z,zx}}{(1 \mu m)^{-1}}\right]\left[\frac{10 GPa}{G}\right] \left[\frac{1 \mu m}{W}\right]^2 \left[\frac{ I_0}{1 amp}\right].
\end{equation}
Here $G$ is the shear modulus of the material and $W$ is the width of the bridge. While we do not know how to estimate $g_{z,zx}$, one might expect that it is of the same order of magnitude as $\bh$.

\section*{Acknowledgements}

We acknowledge financial support through the Simons Investigator Awards by the James Simons Foundation. A. K. was also supported in part by the U.S.\ Department of Energy, Office of Science, Office of High Energy Physics, under Award Number DE-SC0011632. L.R. thanks D. Agterberg and L. Fu for discussions.

\bibliography{Bibliography}

\end{document}